\documentclass[twocolumn,showpacs,preprintnumbers,amsmath,amssymb]{revtex4}
\usepackage{dcolumn}% Align table columns on decimal point
\usepackage{bm}% bold math
\usepackage{epsfig}

\usepackage{color}%

\begin{document}
%\topmargin -1.50cm

%\preprint{APS/123-QED}
\title{Equilibrium properties of the mixed state in superconducting niobium in a transverse magnetic field: Experiment and theoretical model.}% Force line breaks with \\

\author{V. Kozhevnikov$^{1}$, A.-M. Valente-Feliciano$^{2}$, P. J. Curran$^{3}$, G. Richter$^{4}$, A. Volodin$^{5}$, A. Suter$^{6}$ S. Bending$^{3}$ and C. Van Haesendonck$^5$}
\affiliation{
$^1$Tulsa Community College, Tulsa, Oklahoma 74119, USA\\
$^2$Thomas Jefferson National Accelerator Facility, Newport News, VA 23606, USA\\
$^3$University of Bath, Bath BA2 7AY, United Kingdom\\
$^4$Max-Planck-Institut for Intelligent Systems, 70569 Stuttgart, Germany\\
$^5$Solid State Physics and Magnetism Section, KU Leuven, BE-3001 Leuven, Belgium\\
$^6$Paul Scherrer Institut, 5232 Villigen PSI, Switzerland}.\\


\begin{abstract}
\noindent
Equilibrium magnetic properties of the mixed state in type-II superconductors were studied on high-purity film and single-crystal niobium samples with different Ginzburg-Landau parameters in perpendicular and parallel magnetic fields
using dc magnetometry and scanning Hall-probe microscopy. The magnetization curve for samples with unity demagnetizing factor (slabs in perpendicular field) was obtained for the first time. It was found that none of the existing theories is consistent with these new data. To address this problem, a theoretical model is developed and comprehensively validated. The new model describes the mixed state in an averaged limit, i.e. without detailing the samples' magnetic structure and therefore ignoring the surface current and interactions between the structural units (vortices). At low values of the Ginzburg-Landau parameter it converts to the model of Peierls and London for the intermediate state in type-I superconductors. The model quantitatively describes the magnetization curve for the perpendicular field and provides new insights in properties of the mixed state, including properties of individual vortices. In particular, it suggests that description of the vortex matter in superconductors of the transverse geometry as a ``gas-like" system of non-interacting vortices is more appropriate than the frequently used solid-like scenarios.

\end{abstract}\

%\pacs{74.20.-z, 74.25.Ha, 78.70.Nx}% PACS, the Physics and Astro\Deltanomy
                             % Classification Scheme.,

\maketitle INTRODUCTION \vspace{1 mm}

Equilibrium bulk magnetic properties of the mixed state (MS) in type-II superconductors are discussed in all superconductivity textbooks and in numerous papers followed after the discovery of type-II superconductivity by Shubnikov with coworkers eight  decades ago \cite{Shubnikov}. References for many of these papers can be found, e.g., in \cite{Shoenberg,Serin, Brandt, Zeldov}. However, some fundamental magnetic properties of the MS are still not well understood. Examples include, but are not limited to, the magnetization curve $M(H)$, where $M$ is magnetic moment and $H$ is the applied field, and the field strength $H_i$ (also referred to as the magnetic and magnetizing force \cite{Maxwell}, the thermodynamic field \cite{De Gennes}, the Maxwell field \cite{Abrikosov}, \textit{etc.}) in samples of other than cylindrical geometry. As usual \cite{Abrikosov}, under ``cylindrical geometry" we imply infinite right cylinders with base of arbitrary shape (e.g. circular cylinders and slabs) in parallel field, that is samples with demagnetizing factor $\eta=0$ \cite{Landafshitz_II}.

Our original interest to equilibrium properties of type-II superconductors (that is to properties of pinning free type-II samples) of non-cylindrical geometry and specifically to those with $\eta=1$ (infinite slabs in perpendicular  field \cite{Landafshitz_II}), was due to the fact that the latter do not have a lateral surface by definition. This automatically excludes effects associated with surface current (including the Meissner state) and surface barriers.  Hence, properties of sufficiently thick samples of this kind provide access to the pure bulk properties of the MS, i.e. to properties of the vortex assembly. On the other hand, vortices in such samples are evenly distributed due to symmetry and their number density is strictly calculated from the flux conservation. Thus, properties of these samples may also be used for inferring properties of individual vortices. Therefore, knowledge of equilibrium properties of samples of the transverse geometry (this term will be used for samples with $\eta=1$)
is of fundamental importance. % significance.

It is worth noting that real samples of the cylindrical and the transverse geometry are samples with $\eta \rightarrow 0$ and $\eta \rightarrow 1$, respectively. Correspondingly, effects due to the lateral surface are most significant for the former and practically irrelevant for the latter.

Here we report on results of an experimental study of equilibrium (i.e. reversible or thermodynamic) bulk magnetic properties of the MS, measured on  Nb  high-purity film and single-crystal samples \cite{APS}. The field was applied perpendicular and parallel to the samples' plane. To the best of our knowledge such data for the perpendicular field were not reported before. It turned out that none of the existing theories is consistent with these new data. A theoretical model quantitatively accounting for these data is developed and introduced here as well. Similar to the model of Peierls \cite{Peierls} and London \cite{London, London2} for the intermediate state (IS) of type-I superconductors, our model describes the MS in thick samples of any ellipsoidal shape ($0\leq \eta \leq 1$) in an averaged limit, i.e. without detailing the samples' magnetic structure and therefore ignoring the surface current and interactions between vortices; in other words, following the terminology of de Gennes \cite{De Gennes}, it accounts for the properties of the MS in zero-order approximation.  At low values of the Ginzburg-Landau (GL) parameter $\kappa$, our model converts to the model of Peierls and London. We will show that description of the vortex matter in terms of a system of non-interacting vortices is the most appropriate for samples of the transverse geometry.
\vspace{3 mm}

\maketitle PROBLEM STATUS AND MOTIVATION

\vspace{1 mm}

There are two equilibrium states in which superconductors contain domains of normal (N) phase imbedded into superconducting (S) phase.  Those are the IS in type-I and the MS in type-II materials. In the IS, due to positive energy of the S-N interface, the N domains are multi-flux-quantum laminae, whereas in the MS, due to negative interface energy, these domains are single-flux-quantum vortices. This quantitative difference in the flux magnitude results in drastic qualitative differences between properties of type-I and type-II superconductors. In particular, the MS takes place in samples of any shape including the cylindrical one (i.e. for $0\leq \eta \leq 1$), whereas the IS occurs only if $\eta \neq 0$ \cite{Abrikosov57}.
(By ``shapes" we imply shapes of ellipsoids, because only ellipsoids allow rigorous theoretical treatment of the magnetic properties \cite{Maxwell}.)

For this reason significant attention was paid to measurements of magnetic properties of cylindrical samples of
type-II superconductors, which, in particular, led to a fairly good knowledge of the $M(H)$
curve for this geometry \cite{Shubnikov, Livingston, Serin, Finnemore, French1, French2}.

\underline{For cylinders} at $H < H_{c1}$ the Meissner condition $B=0$ allows one to calculate $M(H)$
in three ways: (i) from thermodynamics, (ii) from  magnetostatics (see, e.g. \cite{MM}), or (iii) from  the Maxwell field $H_i$, which in this case equals to the applied field $H$ due to continuity of the tangential component of this field \cite{Tamm}. Then $M$ is calculated from the definition $H_i\equiv B-4\pi m$, where $B$ is the induction and $m$ is the magnetization, which in superconductors is a macroscopic average $M/V$, where $V$ is the sample volume
\cite{Landafshitz_II}.

However, in the MS ($H_{c1}<H<H_{c2}$) complexity of current distribution leaves only one option to calculate
$M$, i.e. through the field $H_i$ via calculation of the average induction $\bar{B}$. Then $4\pi M/V$ is computed as $\bar{B}-H_i$. For cylinders, where $H_i(=H)$ is known, this was done using the GL theory %(valid near the critical temperature $T_c$ \cite{Gor'kov})
near $H_{c1}$ and $H_{c2}$ \cite{Abrikosov57}, and also with use of the London equation in the vicinity of $H_{c1}$ \cite{Abrikosov, L&P}. In these field regions approximate analytical expressions for $\bar{B}$ are available for the extreme type-II limit  $\ln\kappa\gg 1$ \cite{Abrikosov57, Abrikosov, L&P}. At the same time it is supposed that the entire $M(H)$ curve for any $\kappa$ can be calculated via numerical solution of the GL equations \cite{Tinkham}. Note, however, that available approximate  \cite{Abrikosov57, Koppe} and high-precision \cite{Brandt99} solutions of the GL equations are not quite consistent with this supposition. For instance, theoretical $M(H)$ exhibits an infinite slope at $H_{c1}$, whereas the slope of a truly reversible $M(H)$ curve is not infinite  \cite{Finnemore}.

The situation becomes much more complicated for \underline{non-cylindrical} geometry. For the experiments, this is due to a much larger number of pinning centers in the sample area perpendicular to the field. For that reason available data on $M(H)$ for type-II samples of the transverse geometry (see, e.g., \cite{Chang, Miller, Miller-2}) are strongly irreversible and therefore inappropriate
for judgment on thermodynamic properties.

On the theoretical side, the main complication is due to a demagnetizing field $\textbf{H}_d\equiv \textbf{H}-\textbf{H}_i$. The Maxwell field $H_i$ can be rigorously calculated for uniform (meaning that $B$  is homogeneous throughout the sample) ellipsoids \cite{Maxwell, Shoenberg, Landafshitz_II}. If $H$ is parallel to the sample axis, relative to which the demagnetizing factor is $\eta$, then, in CGS units,
\begin{eqnarray}
(1-\eta)H_i+\eta B = H.
\end{eqnarray}
Therefore
\begin{eqnarray}
H_i \equiv H - H_d = H - \eta(B-H_i)=H - \eta\, 4\pi M/V.
\end{eqnarray}

Thus, in uniform samples $H_d=\eta\, 4\pi M/V$. In the Meissner state  the sample is (i) uniform and (ii) $B(=0)$ is known. The former makes possible to calculate $\eta$ (see, e.g., \cite{Landafshitz_II}) and the latter allows one to use  Eq.\,(1) yielding $H_i=H/(1-\eta)$. Then, $4\pi M/V\equiv B-H_i=-H/(1-\eta)$ in full consistency with experiments (see, e.g., \cite{Shoenberg}). However this is not the case for the MS,  where $B$ is not uniform and therefore $\eta$ is not well defined and neither $H_d$, nor $H_i$ are known.

\begin{figure}
\epsfig{file=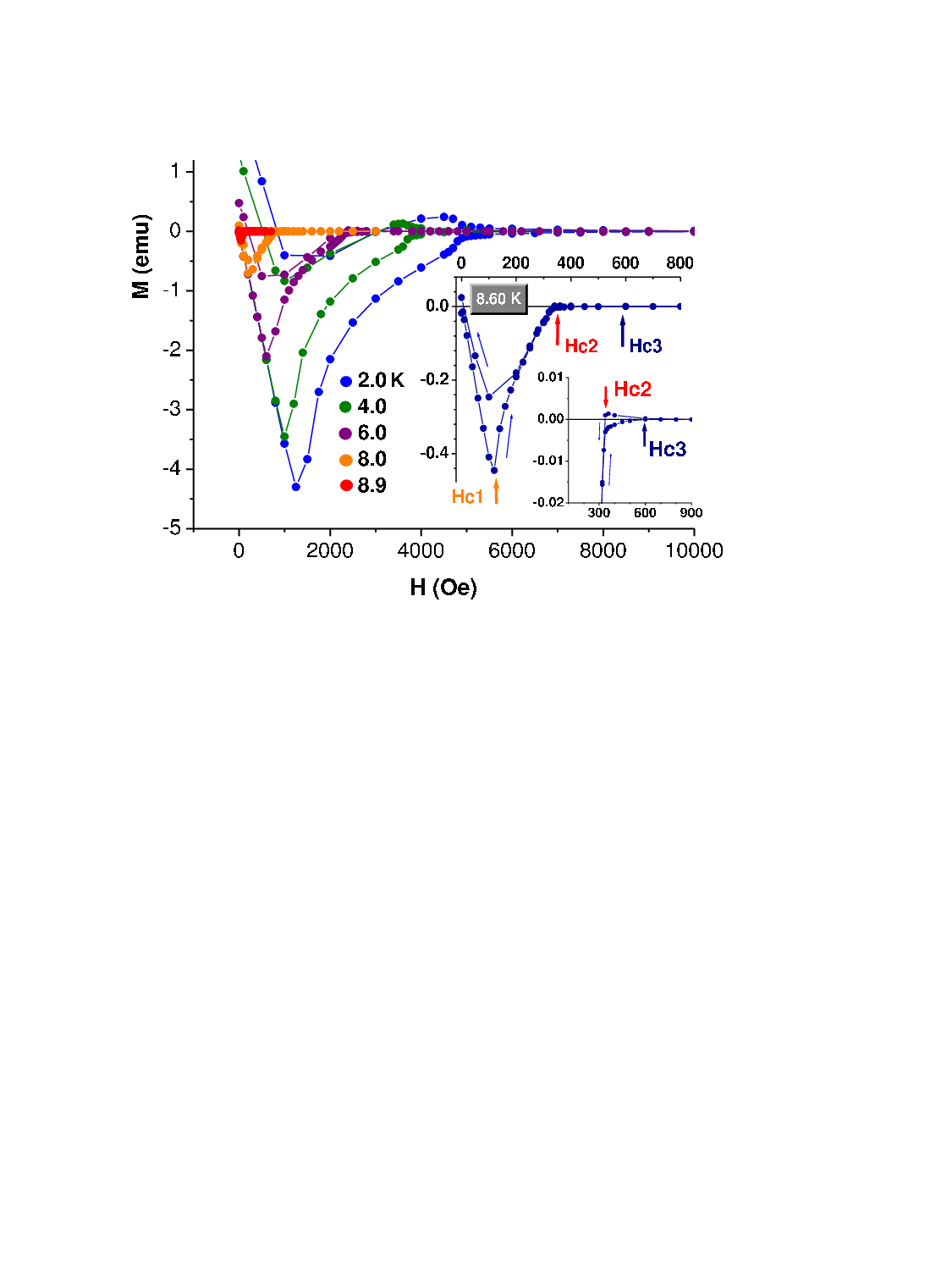,width=6.8 cm}% Here is how to import EPS art
\caption{\label{fig:epsart} Data for the magnetic moment of Nb-SC sample measured in
parallel field at temperatures indicated in the figure. Inserts: $M$ vs $H$ at $T = $ 8.60 K
in two scales. $H_{c1}$, $H_{c2}$ and $H_{c3}$ indicate the respective critical fields.}
\end{figure}

On the first view, solution of the problem for $\eta$
was found long ago by Peierls \cite{Peierls} and London \cite{London} for the IS via replacement of $B$ by $\bar{B}$, which allows one to use $\eta$ of the uniform sample. However, Eq.\,(2), which follows from Eq.\,(1), still contains two unknowns ($H_i$ and $M$) and therefore one more independent relationship is needed to find $M$. Peierls and London resolve this problem \textit{assuming} that $H_i$ equals to thermodynamic critical field $H_c$ in the entire field range of the IS. However, this is not applicable for the MS.

One might object that there is the well known approach for the MS \cite{De Gennes, Fetter} in which $H_i$ for $\eta\neq 0$ is calculated from Eq.~(1) as $H_i=(H-\eta \bar{B})/(1-\eta)$ \cite{comment}. Then  $\bar{B}$ is calculated using $\bar{B}(H)$ obtained for the cylindrical geometry (referred to as the constitutive relation $B_e(H)$ \cite{Fetter}) replacing $H$ by $H_i$, and then both $B_e(H_i)$ and $H_i$ are used to compute $4\pi M/V=B_e(H_i)-H_i$.

 Apart from knowing $B_e(H)$, \textit{the principal  condition} for using this approach is ability to calculate $H_i$ from Eq.\,(1). However, since no new relationship between $H_i$ and $M$ or $B$ is added (see also \cite{Landafshitz_II}), this way to compute $H_i$ is questionable. Indeed, for $\eta = 1$, where $\bar{B}=H$ \cite{Landau-37, Abrikosov, condition}, Eq.~(1) yields $H_i=H(1-\eta)/(1-\eta)=H$ and therefore $M=0$ regardless of $H$, implying that $H_{c2}=\infty$. The reason of such a striking inadequacy of this approach is very simple: a uniform sample with $\eta=1$ is just a sample in the N state, where $M$ is indeed zero. In other words, in order to use Eq.\,(1) for inhomogeneous samples, $H_i$ should be found independently, like in the Peierls-London model for the IS ($H_i=H_c$) or in the cylindrical geometry for the MS ($H_i=H)$ \cite{Abrikosov57}.

In \cite{Brandt_05, Peeters} $M(H)$ for films with different $\kappa$ and thickness $d$ in perpendicular field was calculated using the GL theory. Calculated curves strongly depend on $\kappa$ and $d$, but contradict the rule of 1/2 \cite{MM} (see more about this rule in the Discussion section below) and hence cannot be completely correct. %are irrelevant.
\begin{figure}
\epsfig{file=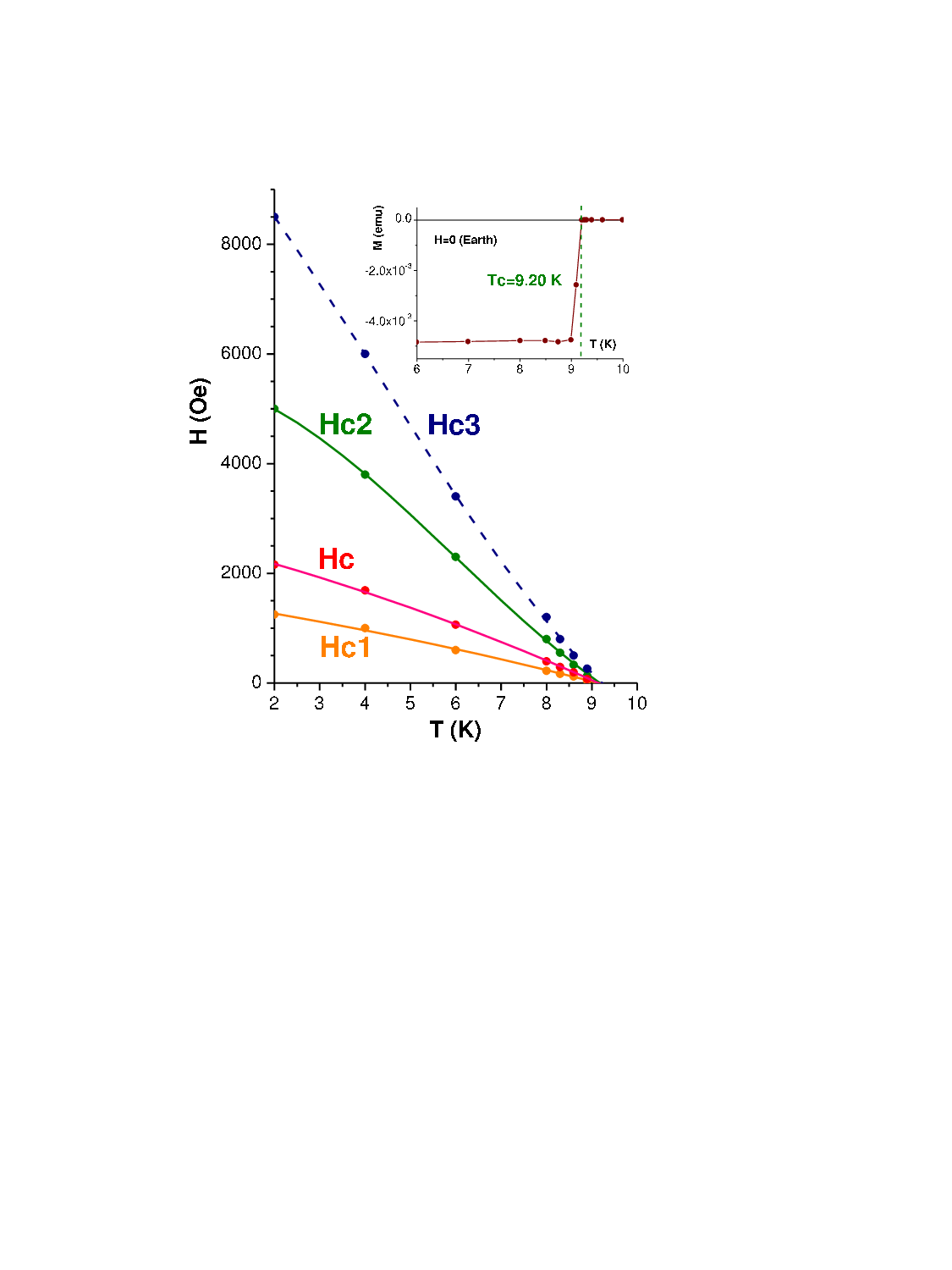,width=6 cm}% Here is how to import EPS art
\caption{\label{fig:epsart} Phase diagram of the single crystal sample Nb-SC measured in parallel field. $H_{c1}$, $H_c$, $H_{c2}$ and $H_{c3}$ indicate the graphs for the respective critical fields and $H_c$ is the thermodynamic critical field. Insert: magnetic moment measured at zero (Earth) field versus temperature.}
\end{figure}

The approach based on London equation \cite{L&P} does not work for the transverse geometry either. Specifically, for cylinders the thermodynamic potential $\tilde{F}\equiv F - BH_i/4\pi = F - BH/4\pi$ is minimized %exclusively
at the expense of the second (negative) term, reflecting the work done by the magnet power supply when the flux through the sample changes. Here $F$ is the Helmholtz potential and $\tilde{F}$ is its Legendre transform, also referred to as the Gibbs free energy \cite{London2}. In the transverse geometry the flux is fixed, hence the term $BH_i/4\pi$ is absent \cite{Abrikosov, De Gennes, Landafshitz_II}. This makes the minimization of $\tilde{F}=F$ impossible, since all terms and their derivatives in $F$ are positive \cite{L&P}.

After all, inhomogeneities of the field and of the vortex cores near the surface perpendicular to the field should be taken into account. These inhomogeneities, unimportant when $\eta=0$, can be important for films in non-parallel fields, making their properties dependent on the film thickness. For instance, in the IS they can change the critical field of a few $\mu$m thick film in perpendicular field by more than 50\% compared to that in parallel field \cite{IS, MM}. An attempt to address this issue for the MS was made by Cody and Miller in experiments with Pb films \cite{Miller-2}, however the results obtained are inconclusive due to strong pinning in their films.

To summarize, (a) available experimental information on the magnetic properties of the MS in non-cylindrical samples is incomplete. In particular, the available $M(H)$ data are strongly irreversible and hence inapplicable %not suitable
for consideration of thermodynamic properties. %and guiding the theory.
(b) Available theoretical results and approaches are controversial. In particular, none of the existing theories is capable to address $M(H)$ curve for samples with $\eta=1$. Progress toward solution of this fundamental problem is the goal of our work, which results are presented below. \vspace{3 mm}
\begin{figure}
\epsfig{file=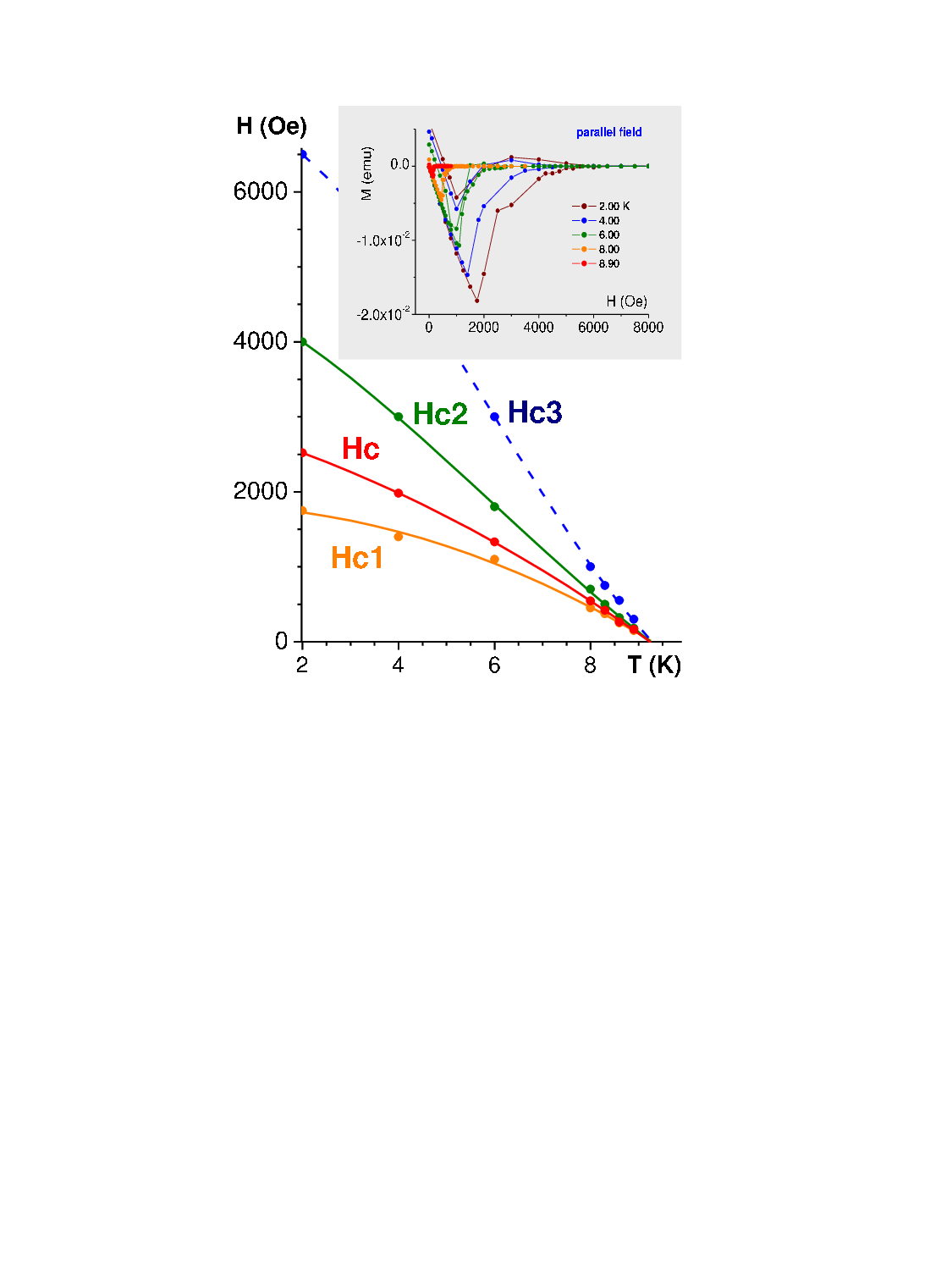,width=4.9 cm}% Here is how to import EPS art
\caption{\label{fig:epsart} Phase diagram of the film sample Nb-F. See captions of Fig.\,2 for notations. Insert: experimental data for the sample magnetic moment in parallel field at indicated temperatures under ascending and
descending field.}
\end{figure}

\maketitle EXPERIMENTAL \vspace{1 mm}

Fabrication of pinning free samples, being very challenging for type-I materials \cite{Shoenberg, MM}, is even more difficult for type-II superconductors, since most of them are alloys with inevitably significant pinning \cite{Shubnikov, De Gennes}. A single crystal sample can be a solution, but since we also need a film for verification of dependencies of the properties on the sample thickness, such a solution is not complete.

Nb is known as a well verified intrinsic type-II superconductor \cite{Finnemore, Khasanov}, hence it is a material from which one can hope to fabricate pinning free films. On this reason Nb was chosen for our samples. However, Nb is also %known as
a getter \cite{Finnemore}. Due to that, our first films, deposited via magnetron sputtering and having residual resistivity ratio RRR up to 70, were still insufficiently clean.

\begin{figure}
\epsfig{file=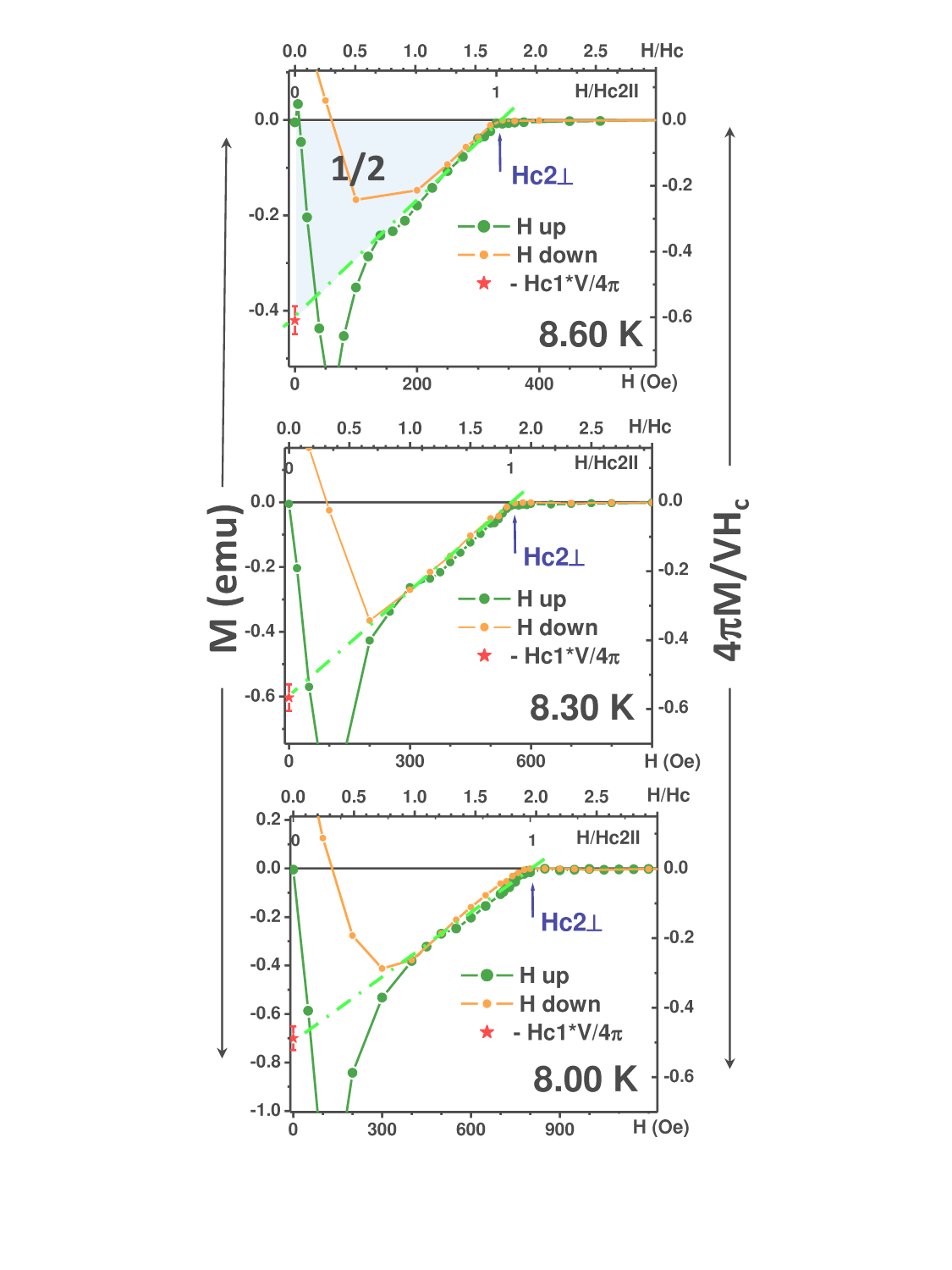,width=6.5 cm}% Here is how to import EPS art
\caption{\label{fig:epsart} Magnetic moment of the Nb-SC sample measured in perpendicular field
at the indicated temperatures. The star shows $M(0) = -H_{c1}V/4\pi$ or $4\pi M(0)/V = -
H_{c1}$. Left and right vertical scales show $M$ in different units as indicated; horizontal
scales give the applied field in Oe (bottom), in units of $H_{c2\parallel}$ (lower top) and in
units of $H_c$ (upper top). 1/2 is the area above the green line in coordinates $4\pi
M/H_cV$ vs $H/H_c$.}
\end{figure}

The Nb film sample (Nb-F)  used in this work is one of two film samples which were used in \cite{filaments}. The film was grown on sapphire via electron cyclotron resonance technique (ECR) \cite{Anne-Marie}; its RRR is 640, the size is 4$\times$6 mm$^2$ and the thickness  is 5.7 $\mu$m. This is a record pure Nb film, of the same purity as In films used in IS studies \cite{IS, MM}. More about ECR grown Nb films can be found in \cite{Anne-Marie 2}.

Another sample (Nb-SC) was also a sample used in \cite{filaments}. It is a one side polished single crystal Nb disc ${\O}$7 mm $\times$ 1 mm supplied by Surface Preparation Laboratory, The Netherlands. %It was a one side polished disc ${\O}$7 mm $\times$ 1 mm.

However, there is one more issue with Nb. A driving force to achieve thermodynamic equilibrium in inhomogeneous samples is the S-N surface tension \cite{MM}, whose magnitude in type-II Nb is significantly less than that in type-I In. This could require even purer Nb samples, but fortunately pinning weakens with temperature $T$. Specifically, $M(H)$ data for both our samples are close to reversible at $T \gtrsim$~8~K, which means that the samples are nearly pinning-free at these temperatures. For this reason, below we mostly discuss the data obtained at high temperatures.
\begin{figure}
\epsfig{file=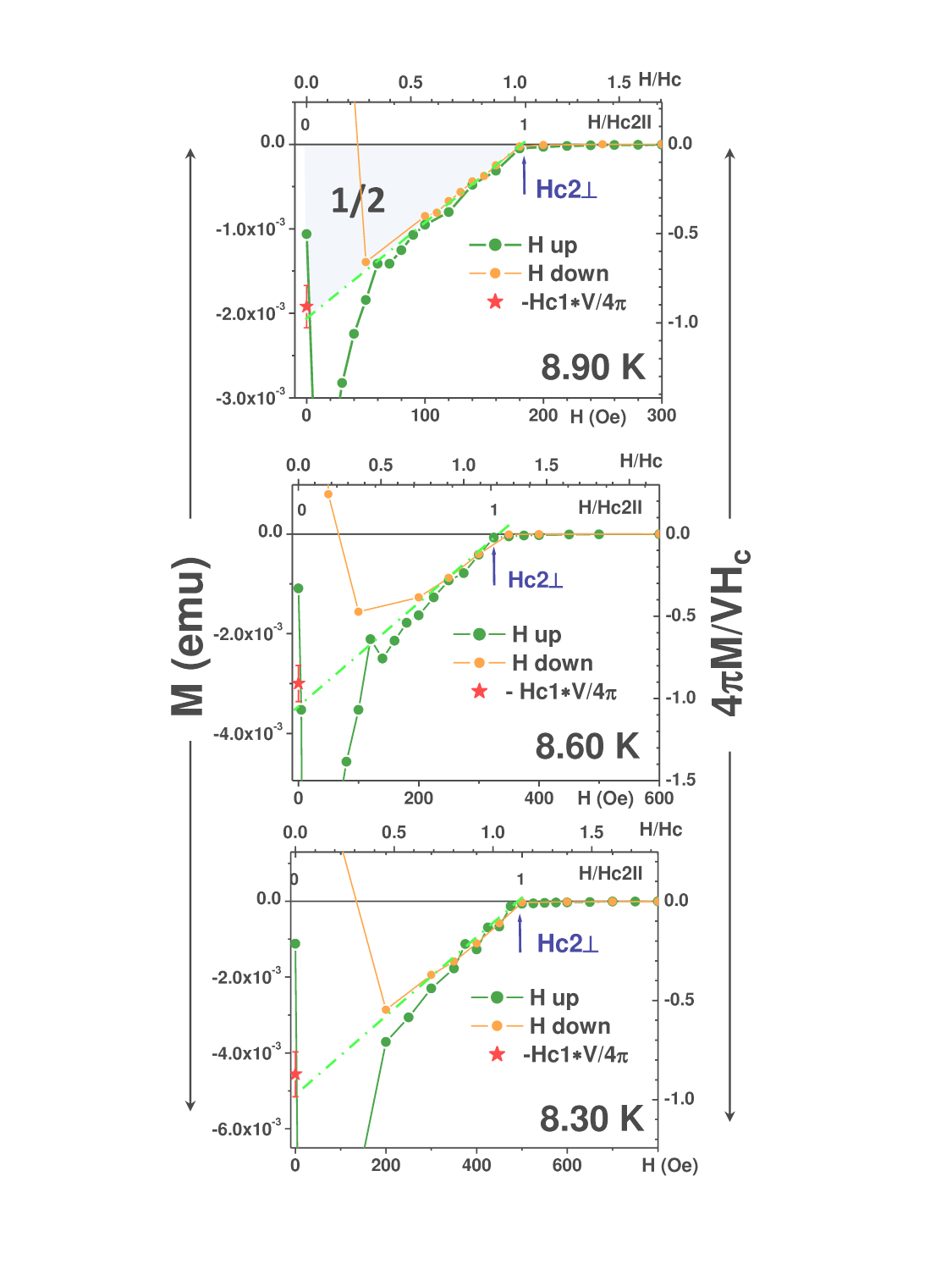,width=6.5 cm}% Here is how to import EPS art
\caption{\label{fig:epsart} Magnetic moment of Nb-F sample measured in perpendicular field
at the indicated temperatures. See caption of Fig.~4 for details.}
\end{figure}

The magnetic moment was measured using Quantum Design $dc$ magnetometer (Magnetic Properties Measurement System).  Data obtained for the Nb-SC sample in parallel field are shown in Fig.\,1 with typical data for high temperatures presented in the inserts. In the Meissner state all data agree with each other. Hence the sample was well aligned with $\textbf{H}$ and the flux trapped at the ascending field was low, which allows calculating the thermodynamic critical field $H_c$ \cite{Livingston}. The sample volume calculated from $M(H)$ at $H<H_{c1}$ agrees with that measured directly within 5\% error, indicating that $\eta_\parallel \lesssim 0.05$ and therefore $\eta_\perp$=1$-2\eta_\parallel \gtrsim 0.9$ \cite{Landafshitz_II}.
Here $\eta_\parallel$ and $\eta_\perp$ are the demagnetizing factors in parallel and perpendicular field, respectively.

The phase diagram of the Nb-SC sample is presented in Fig.\,2 with the data for $M$ vs $T$ at Earth field shown in the insert. The critical temperature $T_c$ of this sample is  9.20 K; $\kappa$ (=$H_{c2}/\sqrt{2} H_c$ \cite{Abrikosov57}) is 1.3 near $T_c$ increasing to 1.6 at 2 K.

The phase diagram of the Nb-F sample is shown in Fig.~3, where the original $M(H)$ data are presented in the insert. The sample volume was determined from the slope of $M(H)$ in the Meissner state; its uncertainty is 10\%. $T_c$ = 9.25 K and $\kappa$ starts from 0.8 near $T_c$ reaching 1.1 at 2 K; $\eta_\perp$ of this sample is 1$- O(10^{-3})$ \cite{Landafshitz_II}.

Data for the magnetic moment measured for the Nb-SC and Nb-F samples in perpendicular field at high temperatures are shown in Figs.~4 and 5, respectively. The data are reversible over more than half of the field range of the MS. Therefore, in this range the sample is in the equilibrium state. The equilibrium $M(H)$ is linear, and its extrapolation (represented by the dash-dotted green line) to $H=0$ yields $4\pi M(0)/V$ close to $-H_{c1}$ (shown by the red star) \cite{M(0)} . The validity of such an extrapolation is supported by the rule of 1/2: the area above the green line equals to the condensation energy $H_c^2V/8\pi$ (=1/2 in coordinates $4\pi M/H_cV$ vs $H/H_c$), where $H_c$ is calculated from the data obtained in parallel field.

\begin{figure}
\epsfig{file=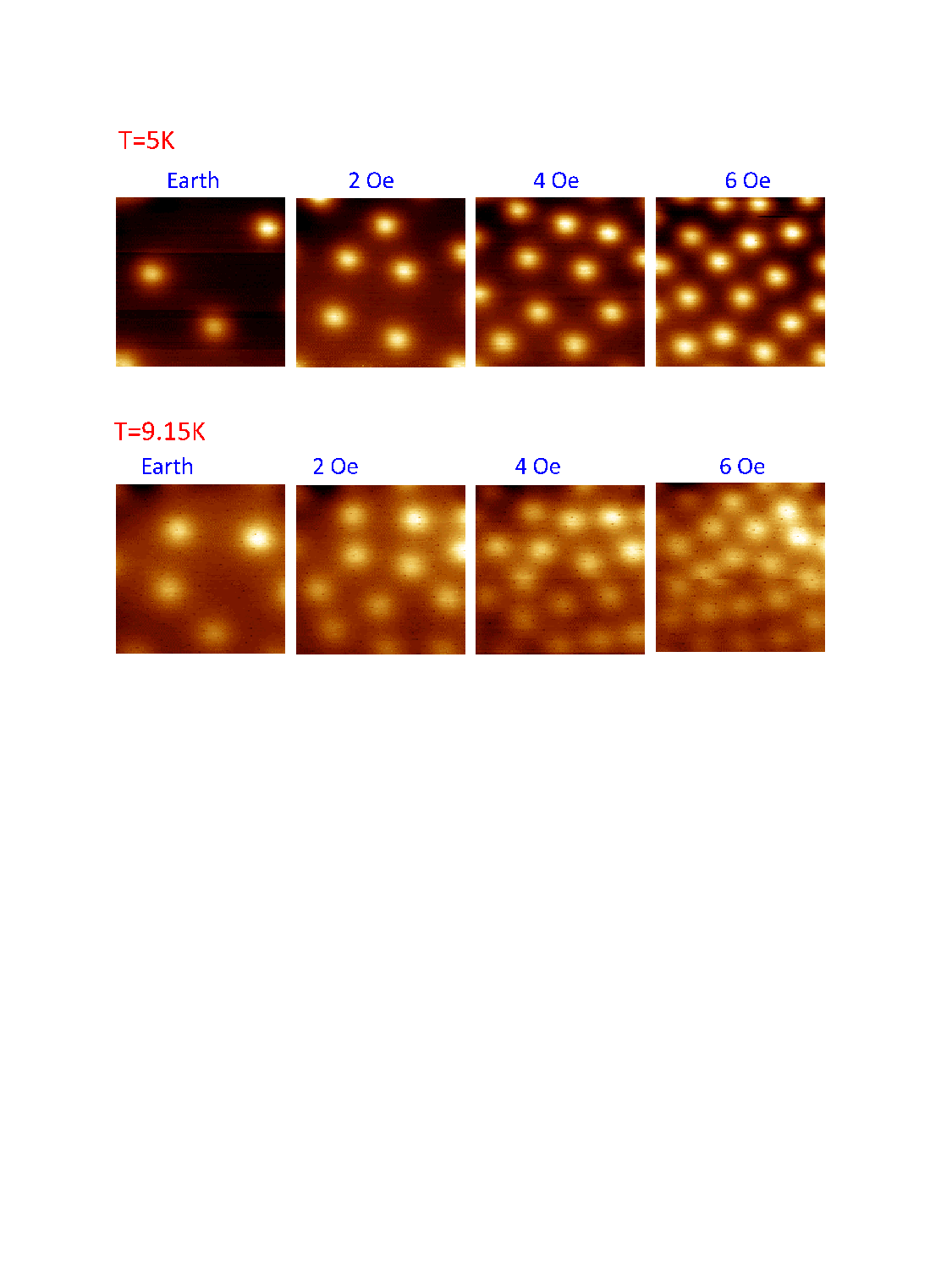,width=8 cm}% Here is how to import EPS art
\caption{\label{fig:epsart} Scanning Hall-probe, images of the MS in Nb film. }
\end{figure}

Comparing Figs.\,4 and 5 with Figs.\,4 and 5 in \cite{MM}, one notices a striking similarity in $M(H)$ for the MS and the IS. However there are also important differences: in the IS $M(0)/V$ and the critical field strongly depend on the sample thickness, whereas for both our samples $4\pi M(0)/V$ is close to $-H_{c1}$ and $H_{c2\perp} = H_{c2\parallel}$, where $H_{c2\bot}$ and $H_{c2\parallel}$ are the critical fields in the perpendicular and the parallel geometry, respectively. Nevertheless, it was important to ensure that our samples are indeed type II superconductors. The most direct way for that is to to measure the flux in the N domains.

In the MS $\bar{B}=n \Phi_0$ \cite{Abrikosov}, where the planar density $n=N/A$ is the number of flux lines $N$ passing through an area $A$. In the transverse geometry $\bar{B}=H$ and therefore
\begin{eqnarray}
n=H/\Phi_0.
\end{eqnarray}
Hence, $n(H)$ allows one to determine the flux in the N domains and therefore the type of superconductor.

With this in mind we probed the film sample with a scanning Hall-probe microscope (SHPM)
\cite{Simon}. The scanned area was 7.6 $\mu$m$\times$7.6 $\mu$m. To achieve better
resolution determined by the contrast of the field inside and outside the flux lines, the
SHPM images were taken at low $H$. At these fields pinning is not small and therefore the
equilibrium hexagonal  structure of the vortex ensemble can be mangled. % damaged.
Typical images taken at $T=$~5.0 K and 9.15 K are shown in Fig.\,6.

A graph for $n-n_0=(N-N_0)/A$ vs $H$ is shown in Fig.\,7. Here $N_0$ is an adjustable
parameter reflecting occasional number of the lines in the scanned area at Earth field due
to pinning and low statistics. As one can see, the experimental points agree with Eq.\,(3),
thus confirming that each flux line carries a single flux quantum $\Phi_0$. Since $\kappa$
of our single-crystal sample is larger than that of the film, we conclude that both our samples are classical type-II superconductors with Abrikosov vortices \cite{Abrikosov57}.

 \vspace{3 mm}

\begin{figure}
\epsfig{file=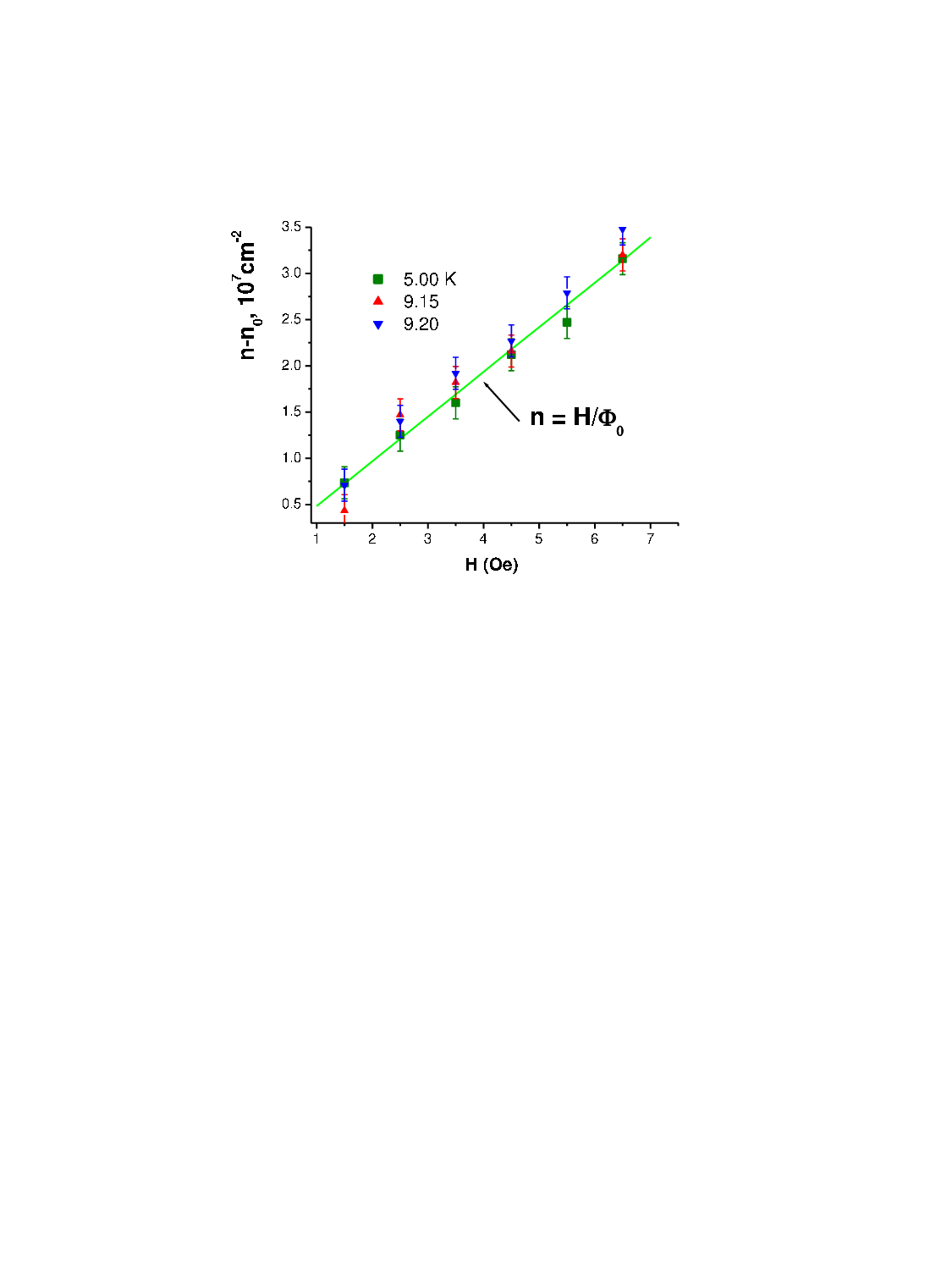,width=4.8 cm}% Here is how to import EPS art
\caption{\label{fig:epsart} Dependence of the flux line density on the applied field at different
temperatures. }
\end{figure}

\maketitle THEORETICAL\vspace{1 mm}

The problem of the magnetic properties of inhomogeneous superconductors was for the first time addressed by Peierls \cite{Peierls} and F. London \cite{London, London2} for the IS. As was mentioned above, both authors solved it in an averaged limit, in which the nonuniform induction $B$ is replaced by averaged $\bar{B}$ and using the demagnetizing factor $\eta$ of the uniform sample. To supplement Eq.\,(1) Peierls and London assumed that $H_i = H_c$. This assumption was justified by a paradigm on instability of the N phase when $H_i < H_c$. Note, however, that this paradigm is valid only for the cylindrical geometry \cite{IS}.

It is important to stress that the averaged description implies that the real sample structure is neglected and therefore any possible interactions between the structural units are neglected automatically. If, e.g., the sample consists of  $N$ unit cells, then the sample free energy $F=N\bar{F}_0$, where $\bar{F}_0$ is the average free energy per unit cell. Therefore, the averaged description does not contain cross terms responsible for interactions, hence excluding them by definition.

The Peierls-London model is valid for thick samples \cite{Shoenberg, IS, MM}, i.e. when the surface related inhomogeneities can be neglected (condition identical to that for the cylindrical geometry). Since N laminae are screened in the sample interior and interact through the outer field, neglect of the near-surface inhomogeneities means neglect of interaction between the laminae, in full consistency with the averaged approximation. Thus the Peierls-London model represents a global description of the IS in a zero-order approximation \cite{De Gennes}, where interaction between the structural units is neglected.

For the MS such an averaged model is missing, resulting in the absence of a global description of this state and leaving a significant ``gap" in understanding the MS magnetic properties. In particular, as shown above, none of the existing theories is capable to address the magnetization curve for a slab in perpendicular field. The model, presented below, is targeted to fill this gap.
\begin{figure}
\epsfig{file=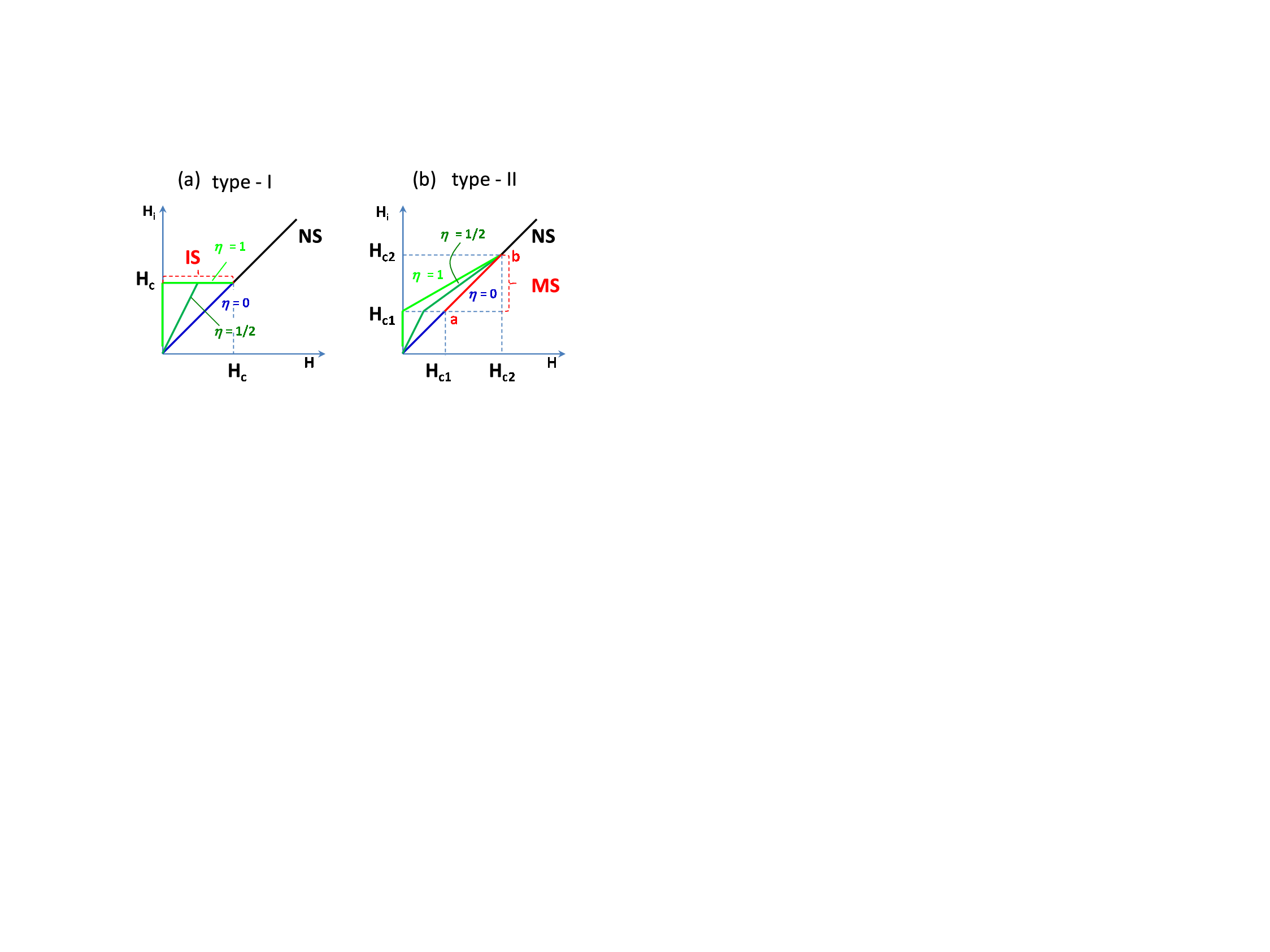,width= 7.5 cm}% Here is how to import EPS art
\caption{\label{fig:epsart} Maxwell field $H_i$ in type-I and type-II superconductors of
different shapes versus applied field $H$. Abbreviations NS, IS and MS designate the normal,
intermediate and mixed states, respectively. Section \textit{ab} represents $H_i(H)$ in the MS for samples with $\eta = 0$. The demagnetizing factors $\eta$ are related to a long cylinder in a parallel field ($\eta = 0$), to a long cylinder in a perpendicular field ($\eta=1/2$) and to an infinite slab in a perpendicular field ($\eta=1$). }
\end{figure}

In Fig.\,8 graphs for $\eta = 0$ represent the Maxwell field $H_i$ vs $H$ for the cylindrical geometry, where $H_i=H$. The red (ab) section in Fig.\,8b represents this dependence for the MS, meaning that the function $H_i(H)$ is linear and extends from $H_{c1}$ to $H_{c2}$. On the other hand, the experimental results for the transverse geometry (see Figs.~4 and 5), specifically (a) linear  $M(H)$, (b) $H_{c2\perp}=H_{c2\parallel}$ and (c) $4\pi M(0)/V = -H_{c1}$, along with the condition $\bar{B}=H$, \textit{directly indicate} \cite{indication} that $H_i(H)$ for $\eta = 1$ is also a linear function extending in the same range. Therefore one can assume a linear form of $H_i(H)$ for all $\eta$, as it is shown in Fig.\,8b. An analytical expression for these functions is
\begin{multline}
H_i= H_{c1}+\frac{H_{c2}-H_{c1}}{H_{c2}-H_{c1}(1-\eta)}[H-H_{c1}(1-\eta)].
\end{multline}
Having $H_i$, one obtains $M$  from Eq.~(2):
\begin{eqnarray}
\frac{4\pi M}{V} =% \frac{H-H^{(i)}}{\eta}=\\
-H_{c1}+\frac{H_{c1}}{H_{c2}-H_{c1}(1-\eta)}[H-H_{c1}(1-\eta)],
\end{eqnarray}
then $\bar{B}=H_i+4\pi M/V$ is
\begin{eqnarray}
\bar{B}=%\frac{1}{\eta}[H-H^{(i)}(1-\eta)]=H^{(i)}+4\pi\frac{M}{V}=\\
\frac{H_{c2}}{H_{c2}-H_{c1}(1-\eta)}[H-H_{c1}(1-\eta)].
\end{eqnarray}
Graphs of these functions are presented in Fig.~9.

In type-I materials, where by definition $H_{c1}=H_{c2}=H_c$ \cite{L&P}, Eq.\,(4) yields $H_i= H_c$, as in the Peierls-London model. This is easily seen from Fig.\,8b: when $H_{c2}\rightarrow H_{c1}$, i.e. when a superconductor converts from type-II to type-I, the graphs in (b) convert to the graphs in (a). Then Eqs.\,(5) and (6) convert to formulas for $M$ and $\bar{B}$ in the Peierls-London model as well (see \cite{MM} for the graphs). Therefore, our model
describes the averaged properties of both the MS and the IS in the limit of non-interacting vortices in type-II superconductors and laminae in type-I superconductors, respectively. \vspace{3 mm}

\maketitle DISCUSSION\vspace{1 mm}

First, we briefly stop at the rule of 1/2 because, being well known (see, e.g.~\cite{Shoenberg}), it is not always clearly articulated in the textbooks. This rule represents the law of energy conservation in superconductors when $\textbf{M}$ is aligned (antiparallel) to $\textbf{H}$. Consistency with this rule is a prerequisite for discussion of equilibrium properties.

In the general case, this law reads that, at constant $T$, the total free energy $\tilde{F}_M$ (defined as $\textbf{M}=-\nabla_\textbf{H}\tilde{F}_M$) of any singly connected superconductor of volume $V$ in a dc magnetic field $\textbf{H}$ of any orientation is
\begin{multline}
\tilde{F}_M(H) =\tilde{F}_M(0)-\int\textbf{M} d\textbf{H}= \\ F_{s0}-\int\textbf{M} d\textbf{H}= F_{n0}-\frac{H_c^2 V}{8\pi}-\int\textbf{M} d\textbf{H},
\end{multline}
where $F_{s0}$ and $F_{n0}$ are the Helmholtz free energies of the S and N states in zero field, respectively.

The first two forms of Eq.\,(7) show that the extra total free energy of a sample (above the free energy of the ground state $\tilde{F}_M(0)= F_{s0}$) is the sample magnetic energy $E_M=-\int\textbf{M} d\textbf{H}$, or the energy of interaction of the applied field with the sample magnetic moment induced by this field. Similar as in conventional diamagnetics  \cite{Tamm}, $E_M$ in superconductors is the kinetic energy of the electrons carrying the induced currents \cite{Shoenberg}. The last form of Eq.\,(7) demonstrates that (a) the total free energy in the N state ($=F_{n0}$) is independent of the applied field since the magnetic permeability $\mu$ of this state is unity, and (b) the source of  $E_M$ is the condensation energy $H_c^2V/8\pi$. Finiteness of the latter makes a transition to the N state a mandatory property of any superconductor \cite{Landafshitz_II}. At this transition $E_M$ of any sample equals to its condensation energy  or area under $M(H)$ curve plotted as $4\pi M/VH_c$ vs $H/H_c$, when $\textbf{M}$ is aligned to $\textbf{H}$, is 1/2. Therefore, the condensation energy density
$H_c^2/8\pi$ is the energy per unit volume it takes to destroy superconductivity i.e. to destroy electron pairing \cite{Kresin}.

As one can see from Fig.~9a, the area under the graphs $M (H)$ for different $\eta$ is the same, meaning that if the magnetic moment is calculated for different orientations of the applied field, the sample condensation energy is the same, as it should. Therefore our model meets the rule of 1/2 and we can proceed to the discussion.

Comparing the modeled magnetization curve for $\eta=1$ in Fig.~9a with the experimental data in Figs.~4 and 5, one can see that for the transverse geometry the model is quantitatively consistent with experiment.

Next, since the area under the graphs in Fig.~9a equals to $H_c^2/2$, we see that $H_c$ is
the geometrical mean of $H_{c1}$ and $H_{c2}$, which is consistent with the rule known for
the extreme type-II limit \cite{Tinkham}. This suggests that the rule $H_{c1}H_{c2}\approx
H_c^2$ is more general than it looked till now.

Further,  as shown by Andreev \cite{Andreev}, in the IS $H_i$ is the Maxwell field and hence the induction in the N domains, where $B=\mu H_i=H_i$, since $\mu=1$ in the N phase. Extending this consideration to the MS \cite{L&P}, one can state that $H_i$ is the Maxwell field in the vortex cores. Therefore our model suggests that $B$  in the vortex cores  increases from $H_{c1}$ at $H=(1-\eta)H_{c1}$ to $H_{c2}$ at $H=H_{c2}$ and the structure of individual vortices (in sufficiently thick samples) does not depend on the sample shape ($\eta$).
\begin{figure}
\epsfig{file=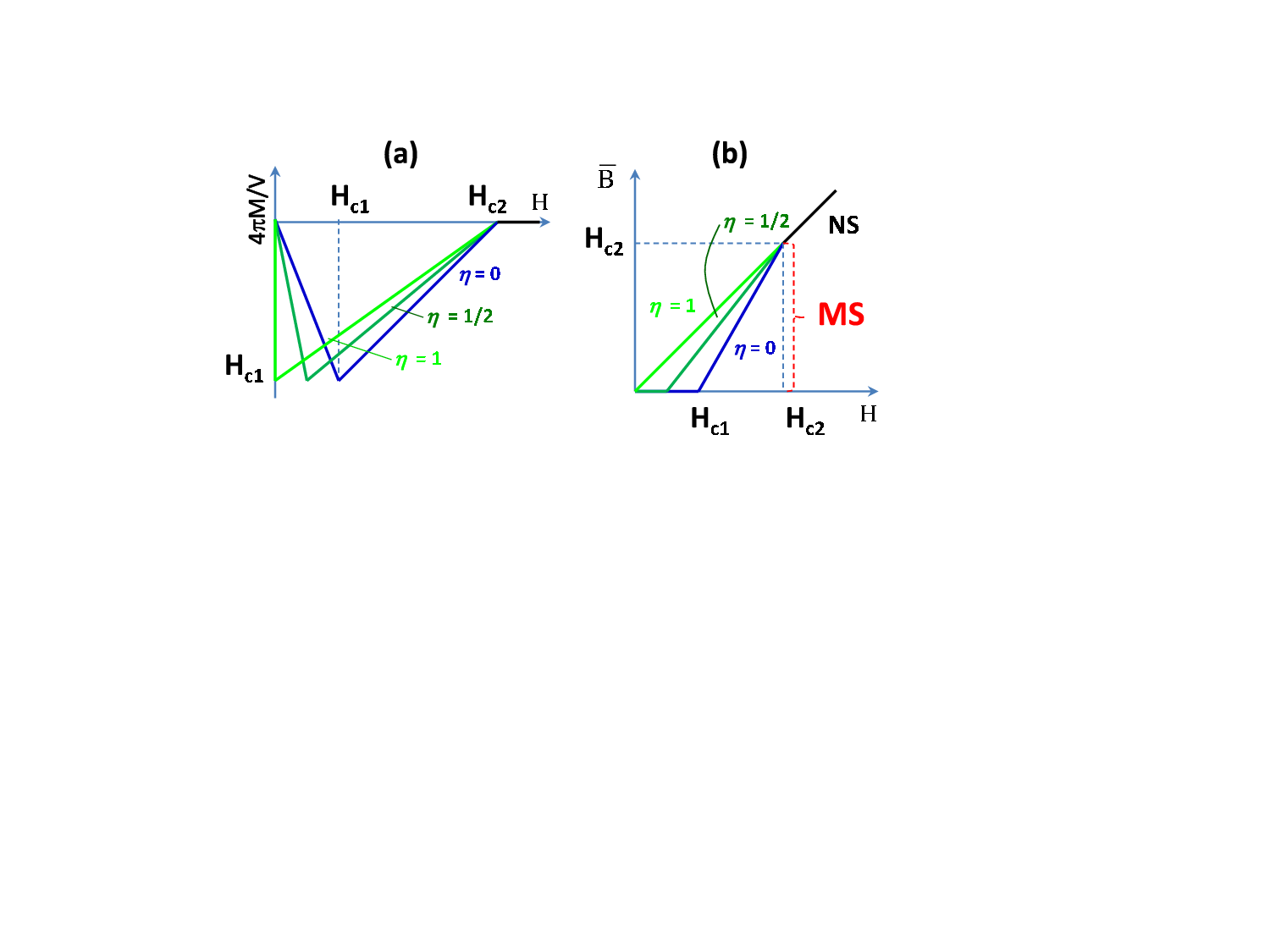,width=8 cm}% Here is how to import EPS art
\caption{\label{fig:epsart} (a) magnetic moment of and (b) averaged induction  in type-II
superconductors with different demagnetizing factors $\eta$ (see caption of Fig. 8 for
explanation). }
\end{figure}

However, there is an obvious problem. In Fig.~9 the linear graph  $M(H)$  for cylinders ($\eta = 0$)
differs from the experimental curve, showing a nonlinear change near $H_{c1}$ (see, e.g., Fig.~1 or \cite{Finnemore}). In the standard theory \cite{De Gennes, Tinkham, L&P} this feature for cylindrical geometry (yielding the infinite slope at $H_{c1}$)
is described assuming overlapping the fields of neighboring vortices, leading to their repulsion.  On the other hand, in cylinders there is the surface current, which is absent in samples of the transverse geometry. The magnitude of this current is determined by the field drop at the surface \cite{Landafshitz_II}, which is maximal near $H_{c1}$ and vanishes upon approaching $H_{c2}$. Our model includes neither the vortex-vortex interaction, nor surface current, which explains the difference of the modeled and experimental $M(H)$ curves for $\eta = 0$. Note that surface current also is \textit{not} accounted for in the standard theory \cite{De Gennes, Tinkham, L&P}. %Interesting that the same feature in the $M(H)$ curve (the infinite slope near $H_{c1}$ for  cylindrical geometry) was also obtained in a laminae model of the MS \cite{Gorter}. % at the expense of a ``rather arbitrary form" of free energy of a boundary layer between the laminae

The question then arises, whether such a model is needed, if vortices interact? The same question can be formulated as: Why for samples of the transverse geometry the experimental $M(H)$ graphs in Figs.~4 and 5 are identical to that in the model of noninteracting vortices in Fig.~9a? The only possible answer we can suggest for both formulations is: Because vortices in samples of the transverse geometry do not interact.

The field passes through any superconducting sample of the transverse geometry in which currents are optimized so to minimize the sample magnetic energy and therefore the total free energy $\tilde{F}_M$. However, the result of this optimization is different in type-I and type-II superconductors. Type-I samples tune the period of the laminar structure, the fraction of the normal phase and induction inside it, as well as the currents and the laminae shape near
the surfaces, all together leading to a strong dependence of the magnetic properties on the
sample thickness $d$ \cite{IS, MM}.

No doubt that all these degrees of freedom are also available for type-II superconductors,
but owing to the gain received from the negative interface energy, in this case all tunings
are performed keeping maximum possible number of vortices and therefore the minimum flux passing
through them, i.e. $\Phi_0$. Then the vortex density $n$ is as that in Eq.~(3) and the parameter
of the most effectively packed hexagonal vortex lattice $b=(2\Phi_0/\sqrt{3}H)^{1/2}$.

It is important to stress that $n$ and $b$ are as such due to the symmetry and the flux conservation and, hence, do not depend on $d$.

On the other hand, if vortices in the transverse geometry do not interact, the sample magnetic energy is a simple sum of the free energies of the individual flux lines $\epsilon d$, where $\epsilon$ is the energy of the flux line per unit length (line tension). Therefore, considering that $H=\bar{B}=n\Phi_0=N\Phi_0/A$ and using Eq.\,(5) with $\eta=1$ one obtains
\begin{multline}
E_M=-\int\textbf{M}d\textbf{H}=N\epsilon d=\frac{V}{4\pi}H(H_{c1}-\frac{H_{c1}}{2H_{c2}}H)=\\
Nd\frac{\Phi_0}{4\pi}(H_{c1}-\frac{H_{c1}}{2H_{c2}}H),
\end{multline}
which yields
\begin{eqnarray}
\epsilon=\frac{\Phi_0 H_{c1}}{4\pi}(1-\frac{H}{2H_{c2}}).
\end{eqnarray}

Comparing $\epsilon$ at $H_{c1}$ in the cylinders ($=\Phi_0 H_{c1}/4\pi$ \cite{De Gennes,Tinkham,Abrikosov,L&P}) with Eq.\,(9), we see that $\epsilon$ in the transverse geometry at $H\rightarrow 0$ when $H_i=H_{c1}$ equals to $\epsilon$ in the cylinders, as \textit{it should}. At higher field, $\epsilon$ decreases becoming $\Phi_0H_{c1}/8\pi$ at $H$=$H_{c2}$, where it yields
\begin{multline}
E_M=N\epsilon d=
(\frac{H_{c2}A}{\Phi_0})(\frac{\Phi_0H_{c1}}{8\pi})d=\frac{H_{c1}H_{c2}}{8\pi}V=\frac{H_c^2}{8\pi}V,
\end{multline}
as \textit{it should} as well.

 This confirms  the validity of Eq.\,(9) and, hence, the absence of the interaction between vortices in samples with $\eta=1$.

Therefore, the vortex matter in samples of the transverse geometry can be viewed as a peculiar 2D ``gas-like" system  of non-interacting vortices where the role of pressure $P$ is taken by the applied field $H$ and the equation of state is given by Eq.~(3).

Indeed, similar to the gas,  whose density $\rho \rightarrow 0$ when pressure $P\rightarrow0$, and isothermal compressibility $\rho^{-1}(\partial \rho/\partial P)_T=1/P$, the vortex matter is highly compressible,  i.e. its density $n\rightarrow0$ at $H\rightarrow0$ and the ``compressibility" $n^{-1}(\partial n/\partial H)_T=1/H$. Similar to the gas, where the slope of $\rho$ vs $P$ is determined by the Boltzmann constant, in the vortex matter the slope of $n$ vs $H$ is determined by the fundamental constant $\Phi_0$. But, contrary to the gas, properties of the vortex matter do not depend on temperature. When the field $H$ (``pressure") changes, the vortex density $n$ changes accordingly. But when the field is fixed, vortices do not move. Therefore, vortices in a slab in perpendicular field can be treated as a system at zero temperature. But zero temperature means zero entropy. Therefore vortices should be ordered in full consistency with well known experimental fact \cite{Essmann}.

Comparing the gas-like scenario with solid-like pictures \cite{Larkin}, one can see that the former is more appropriate for the vortex matter since a primary property of solids, rigidity, is absent in the equilibrium vortex ensemble.

Now we turn to one more aspect of our experimental results, i.e. the %the aspect associated with
proximity of $\kappa$ of our film sample to the critical value $\kappa_c= 1/\sqrt{2}$. In recent years there emerged an active interest in properties of superconductors with $\kappa \approx \kappa_c$ (see \cite{Milosevic} and references therein). Considering properties of the critical ($\kappa =\kappa_c$) superconductors in the framework of classical field theory, Bogomol'nyi  showed \cite{Bogomolnyi} that vortices with flux exceeding a single flux quantum are unstable against decay to single flux quantum vortices, and that the energy of a system of stable (single-flux-quantum) vortices equals to the sum of the energies of the unit vortex, i.e. vortices do not interact.

The GL parameter of our film sample is 1.1$\kappa_c$, and, as we see, the vortices are indeed single-flux-quantum non-interacting units. The same was found for the Nb-SC sample with $\kappa=1.8\kappa_c$.
Hence, our experimental results confirm Bogomol'nyi's predictions for superconductors with $\kappa \gtrsim 1/\sqrt{2}$.

Finally, we have to address a question inevitably arising when reading this paper: if vortices do not interact when the sample (e.g. a film) is in perpendicular field, why do they interact when the field is parallel, as stated in the standard theory? In view of enormous amount of literature associated with the vortex interaction, a complete answer to this question is hardly feasible in the framework of this paper. Due to that  we will limit ourselves to reminder that  the vortex-vortex interaction in the standard theory follows from the \textit{assumption} of overlapping of the inductions $B$ of neighboring vortices, which increases the sample free energy thus  leading to repulsive interaction between vortices \cite{De Gennes, Tinkham, L&P}. Note, however, that overlapping of $B$ at some point inside the sample means overlapping of currents at the same point, i.e. overlapping of vortices \cite{overlapping}. The latter is met neither in normal fluids \cite{GL_Hydrodynamics} nor in superfluids \cite{Feynman, Yarmchuk}. In the GL theory, vortices in superconductors also do not overlap \cite{Abrikosov57, Brandt2003}. %Therefore the assumption of the $B$ overlapping and hence the standard theoretical explanation of the vortex-vortex interaction may be questionable.

Indeed, in regular matter overlapping of electron shells of neighboring molecules  results in strong molecular repulsion leading to practical incompressibility of liquids and solids. As we have seen, this is not the case for superconductors. This is because molecules are fixed entities, whereas vortices are self-adjustable units. Recall that currents induced in a singly connected superconductor serve \textit{solely} to reduce its free energy. Therefore, since the $B$ overlapping increases the free energy, one can expect that superconductor will tune the currents to avoid that, thus keeping vortices non-interacting regardless of the vortex density. This qualitative scenario is consistent with the reported experimental results for the transverse geometry and, most importantly, with the rule of 1/2, valid for all geometries and indicating that the total free energy of superconductors contains no potential energy. %to the absence of potential energy in the total free energy of superconductors.

Coming back to our  model of zero-order approximation, we note that it does not include inhomogeneities near the "transverse"  (perpendicular to the field) sample surface, like in the Peierls-London model for the IS. These inhomogeneities increase the sample magnetic energy and therefore modify the magnetization curve. In particular, as it was mentioned above, in the IS they can significantly decrease the upper critical field \cite{MM}. However, contrary to the IS, where effects of these inhomogeneities were noticed already in 2 mm thick samples \cite{Sharvin-Sn}, such effects were not found in our samples. This can be explained by a finer pattern of these inhomogeneities (compare images in Fig.~6 with those in \cite{IS}). Therefore, the surface related effects in the MS should be expected in thinner samples and they may differ from those in the IS, hence constituting a very intersting problem of fundamental superconductivity.

\vspace{3 mm}

\maketitle SUMMARY AND OUTLOOK\vspace{1 mm}

Equilibrium properties of the mixed state in type-II superconductors were studied with high-purity film  and
single-crystalline  niobium samples with zero and unity demagnetizing factor $\eta$, that is in parallel and perpendicular magnetic field, respectively. The magnetization curve for the samples with $\eta$=1 was obtained for the first time. It was found that existing theories fail to describe these new data. A theoretical model successfully addressing this problem was developed and experimentally validated.

The new model describes magnetic properties of the mixed state in a zero order approximation where
interactions between vortices and surface current are ignored. The model is applicable to thick samples with any $\eta$ without limitation for the magnitude of the Ginzburg-Landau parameter $\kappa$. The model is quantitatively consistent with the data obtained for the samples with $\eta=1$, where the surface current is absent by definition. This indicates to the absence of interaction between vortices in such samples. An expression for the field strength $H_i$ inside superconductors in the mixed state is obtained  together with a formula for the line tension of vortices valid in the entire field range of the mixed state.  %It is shown that the root mean square value of the induction at an applied field $H\rightarrow H_{c2}$ is the thermodynamic critical field $H_c$.
At low $\kappa$ our model converts to that of Peierls and London for the intermediate state in type-I superconductors, which is valid in the limit of non-interacting laminae. It is shown that visualization of the vortex matter as an ordered 2D ``gas-like" system at zero temperature is more appropriate than the frequently used solid-like scenarios.

%It is found that effects of the surface related inhomogeneities differ from those in the intermediate state in type-I superconductors, but the specific form of these effects remains to be revealed. Investigation of these effects is important for further progressing in understanding of properties of type-II superconductors.

The reported model is constructed and verified using experimental results obtained with low-$\kappa$ Nb type-II superconductors. Therefore it is interesting and important to test the model with materials of higher $\kappa$. Single-crystal samples of A15 compounds, e.g. V$_3$Si, can be appropriate candidates for such a verification. Single crystal samples of unconventional superconductors close to the critical temperature could be interesting  as well.

\vspace{4 mm}

\maketitle Acknowledgments \vspace{1 mm}

We express a deep gratitude to Michael E. Fisher and Konstantin A. Kikoin for valuable remarks and to Yurii M. Kagan for encoraging comments on the manuscript. We are thankful to Oscar Bernal, Connie Hebert and Andrew MacFarlane for their crucial help in organizing the project.

This work was supported in part by the National Science Foundation (Grant No. DMR 0904157), by the Research Foundation -- Flanders (FWO, Belgium) and by the Flemish Concerted Research Action (BOF KU Leuven, GOA/14/007) research program. V.K. acknowledges support from the sabbatical fund of the Tulsa Community College.  A.-M. Valente-Feliciano is supported by the U.S. Department of Energy, Office of Science, Office of Nuclear Physics under contract DE-AC05-06OR23177.% \vspace{3mm}



\end{document}